\documentclass[a4paper,12pt]{article}

\usepackage[tbtags]{amsmath}

\newcommand{\ud}{\mathrm{d}}
\renewcommand{\vec}{\mathbf}

\title{\bf New approach to nonrelativistic ideal magnetohydrodynamics}

\author{{\normalsize \textbf{Rabin Banerjee$^{1}$ and Kuldeep Kumar$^{1,2}$}}\\[0.0ex]
{\small $^1$\textit{S.N. Bose National Centre for Basic Sciences,}}\\[-0.8ex]
{\small \textit{JD Block, Sector III, Salt Lake, Kolkata 700098, India}}\\[-0.4ex]
{\small $^2$\textit{Department of Physics, Panjab University, Chandigarh 160014, India}}\\[-0.4ex]
{\small \texttt{rabin@bose.res.in}, \texttt{kuldeepk@pu.ac.in}}
}

\date{}

\begin{document}

\maketitle

\begin{abstract}
We provide a novel action principle for nonrelativistic ideal magnetohydrodynamics in the Eulerian scheme exploiting a Clebsch-type parametrisation. Both Lagrangian and Hamiltonian formulations have been considered. Within the Hamiltonian framework, two complementary approaches have been discussed using Dirac's constraint analysis. In one case the Hamiltonian is canonical involving only physical variables but the brackets have a noncanonical structure, while the other retains the canonical structure of brackets by enlarging the phase space. The special case of incompressible magnetohydrodynamics is also considered where, again, both the approaches are discussed in the Hamiltonian framework. The conservation of the stress tensor reveals interesting aspects of the theory.
\end{abstract}

%%%%%%%%%%%%%%%%%%%%%%%%%%%%%%%%%%%%%%%%%%%%%%%%%%%%%%%%%%%%%%%%%%

\section{Introduction}

Understanding a system from an action principle is always desirable as it offers further insights. The action principle for ideal magnetohydrodynamics (MHD) was originally proposed by Newcomb \cite{Newcomb1962}, both in Euler and Lagrange variables, and followed by others \cite{Lundgren1963, Calkin1963, PH1966, Merches1969, BO2000, ZQBB2014, Y2015}. However, the form of the Lagrangian is not unique but varies from author to author, who have employed different approaches, and also in the number of basic fields in the Lagrangian. The roots of these ambiguities lie in fluid dynamics itself \cite{Jackiw2004, BGM2015}.

Writing down a Hamiltonian for a given system is reasonably straightforward as its form can be written on general principles. Appropriate brackets may be suitably defined to yield known equations of motion. In most cases this is easier said than done. This is because these brackets have to satisfy several properties, like antisymmetry, distributiveness and associativity. The last, which is encoded in the Jacobi identity, is quite difficult. Despite these obstacles, nevertheless, a Hamiltonian formulation for ideal MHD in terms of physical fields---fluid density, entropy density, fluid velocity and magnetic field---was given in \cite{MG1980}, where a general form of noncanonical bracket was posited. An algorithm for writing such a noncanonical bracket was elaborated in \cite{Morrison1982}. One has to choose a suitable form of velocity in terms of Clebsch variables \cite{Morrison1982, Zakharov1971, Morrison1998} and identify the canonical pairs in the Hamiltonian. The usual (canonical) Poisson bracket can then be mapped using the chain rule of functional derivatives to the noncanonical Poisson bracket of \cite{MG1980} involving physical fields. One can use a Clebsch-type decomposition for magnetic field as well. An alternative constructive approach from the Lagrange-variable description has been detailed in \cite{Morrison2009}. In this paper we present another approach to obtain the noncanonical brackets starting from an action principle and following Dirac's constraint analysis \cite{Dirac1964}. It is worthwhile to mention here that the use of Dirac brackets for reduction of the general (compressible) MHD to incompressible MHD already exists in the literature \cite{CMT2012, CGBTM2013, Chandre2015}, but the use of Dirac brackets presented here to obtain the noncanonical brackets of general (compressible) MHD is completely new.

The MHD Lagrangian proposed in \cite{BO2000} uses density, entropy density, velocity, magnetic field and a new field subject to a constraint, introduced by Lin \cite{Lin1963}, as the basic fields. The continuity equation, entropy equation and Faraday's law are incorporated in the Lagrangian as constraints along with the Lin's constraint by augmenting the basic fields with Lagrange multiplier fields necessary to enforce these constraints. The Gauss's law for magnetism is not incorporated in the Lagrangian itself but is later used to correctly reproduce the force-balance equation, also commonly known as the Euler equation. In this approach, however, the physical significance of Lin's constraint remained obscure. Another approach is discussed in \cite{Y2015}, where variations of the action with respect to velocity and magnetic field yield a Clebsch-type representation for these variables.

A new approach to obtain the Lagrangian for nonrelativistic perfect fluids, based on Noether's definition of energy-momentum tensor, was advocated in \cite{BM2015}, a paper involving one of us. This approach naturally dictates a Clebsch-type parametrisation of velocity. The ambiguities of introducing by hand the Lin's constraint are thereby avoided.

In this paper, as an extension of this approach \cite{BM2015}, we propose a Lagrangian for ideal nonrelativistic MHD from which the equations of motion are reproduced. The basic fields in this Lagrangian are the fluid density, entropy density, magnetic field and some additional fields. Also, a generalised Clebsch form of the velocity involving the magnetic field is obtained, which is similar to the one considered in \cite{Morrison1982}. Then we discuss a systematic passage to the Hamiltonian formulation using Dirac's constraint analysis \cite{Dirac1964}. The MHD system, in Dirac's classification, turns out to be second-class. Thus all constraints may be eliminated by using Dirac brackets instead of Poisson brackets. These Dirac brackets are just the noncanonical brackets first posited in \cite{MG1980}. The Hamiltonian therefore has the standard canonical structure from which the MHD equations are reproduced by using the Dirac (noncanonical) brackets. However, we also give a modified Hamiltonian from which the MHD equations can be deduced using the canonical brackets. This is done by enlarging the phase space so that the Hamiltonian no longer has its earlier canonical form. Following the same approach based on Dirac's algorithm, we also consider the case of incompressible MHD. Again, one can opt either for noncanonical brackets and the standard Hamiltonian, or for the standard canonical brackets and a modified Hamiltonian. The compatibility of both approaches is shown and a comparison with existing results \cite{CMT2012} has also been done. Finally, we discuss the conservation of the stress tensor.

In Sec.~\ref{mhdrev} we present a brief review of ideal MHD which also helps to set up notations. An action principle is given in Sec.~\ref{mhdact} where we rederive the MHD equations using the variational approach. An essential ingredient is the abstraction of a generalised Clebsch decomposition of the velocity that involves the magnetic field. Section~\ref{mhdham} deals with the Hamiltonian formulation in two descriptions. In the first part we discuss the emergence of noncanonical brackets as Dirac brackets while the Hamiltonian retains its canonical structure. The second part, which is complementary to the first, uses an enlarged phase space. The Hamiltonian changes from its canonical form but all the brackets are canonical. Once again the MHD equations are reproduced. In addition to some consistency checks, the role of Gauss's law is also highlighted. Section \ref{mhdinc} is devoted to the Hamiltonian formulation of incompressible MHD where again we provide two equivalent descriptions, one in terms of noncanonical brackets and the other in terms of canonical brackets. We carry out an explicit computation of the Noether's energy-momentum tensor in Sec.~\ref{mhdemt} and demonstrate its conservation. Finally we summarise our results and discuss some future prospects in Sec.~\ref{conclu}.

%%%%%%%%%%%%%%%%%%%%%%%%%%%%%%%%%%%%%%%%%%%%%%%%%%%%%%%%%%%%%%%%%%

\section{\label{mhdrev}Brief review of MHD}

MHD concerns itself with the study of low-frequency interaction between electrically conducting fluids and electromagnetic fields. In the ideal case the flow is nondissipative and the fluid has infinite conductivity. There are two forms of MHD, one in terms of Lagrange variables and the other in terms of Euler variables.  We shall work in the Eulerian form, which is akin to the classical field theory in physical spacetime. Confining to the nonrelativistic case, the basic equations of ideal MHD read\footnote{Notation: $\partial_0=\partial/\partial t$, $\partial_i=\partial/\partial x_i\; i=1,2,3$, summation over repeated index implied, SI units for electrodynamics.}
\begin{gather}
\label{eqrho} \partial_0 \rho + \vec{\nabla} \cdot (\rho \vec{v}) = 0, \\
\label{eqs} \partial_0 s + \vec{v} \cdot \vec{\nabla} s = 0, \\
\label{eqB} \partial_0 \vec{B} = \vec{\nabla} \times (\vec{v} \times \vec{B}), \\
\label{eqv} \partial_0 \vec{v} + (\vec{v}\cdot \vec{\nabla}) \vec{v} = -\frac{1}{\rho} \vec{\nabla} p + \frac{1}{\mu \rho} (\vec{\nabla}\times \vec{B}) \times \vec{B}.
\end{gather}
Equation \eqref{eqrho} is the continuity equation, obeyed by fluids, which expresses the conservation of matter, $\rho(\vec{x},t)$ being the density and $\vec{v}(\vec{x},t)$ the velocity of the fluid. Since there is no dissipation, the fluid motion is adiabatic and the entropy per unit mass $s(\vec{x},t)$ must be conserved along the flow---this is Eq.~\eqref{eqs}. Infinite conductivity implies $\vec{E} + \vec{v} \times \vec{B} = 0$, which can be used to eliminate $\vec{E}$ in Faraday's equation, $\partial_0 \vec{B} + \vec{\nabla} \times \vec{E} = 0$, yielding Eq.~\eqref{eqB}.

The low-frequency version (neglecting displacement current) of Amp\`ere's law, $\mu \vec{J} = \vec{\nabla}\times \vec{B}$, can be used to eliminate $\vec{J}$ to write the Lorentz force on a volume element $\ud V$ as $[\vec{J}\times \vec{B}]\,\ud V = (1/\mu)[(\vec{\nabla}\times \vec{B})\times \vec{B}]\ud V$. In addition to this force, there is also a force on the volume element due to the fluid pressure $p(\vec{x},t)$, which is given by $-(\vec{\nabla}p)\ud V$. Equating the total force to the product of mass, $\rho\, \ud V$, and acceleration, $\ud \vec{v}/\ud t = \partial_0 \vec{v}+ (\vec{v}\cdot \vec{\nabla})\vec{v}$, gives the MHD Euler equation, Eq.~\eqref{eqv}. Apart from Eqs.~\eqref{eqrho}--\eqref{eqv}, we also have Gauss's law for magnetism,
\begin{equation}
\label{gaussB} \vec{\nabla} \cdot \vec{B} = 0.
\end{equation}
The effect of gravity on the fluid motion has been ignored.

%%%%%%%%%%%%%%%%%%%%%%%%%%%%%%%%%%%%%%%%%%%%%%%%%%%%%%%%%%%%%%%%%%

\section{\label{mhdact}Action principle for MHD}

Obtaining equations of motion for MHD, as done in the previous section, is not a cumbersome task. However, as pointed our earlier, a systematic formulation of MHD in terms of an action principle is not as straightforward. In this paper we shall follow the approach discussed in \cite{BM2015}. Let us very briefly recall the gist of this approach. For the simplest case of isentropic ideal fluids (constant $s$), the Hamiltonian (density) is
\begin{equation}\label{ham-fd0}
\mathcal{H} =  \frac{1}{2} \rho v^2 + \rho \epsilon(\rho),
\end{equation}
$\epsilon$ being the thermodynamic internal energy per unit mass. A new field $\theta$, conjugate to $\rho$, is introduced, taking the velocity as dependent on $\theta$, so that $\rho$ and $\theta$ are the basic fields. This requires us to write the Lagrangian (density) as 
\begin{equation} \label{lang-fd00}
\mathcal{L} =  \rho \partial_0 \theta  - \Big( \frac{1}{2} \rho v^2(\theta) + \rho \epsilon(\rho) \Big).
\end{equation}
We recall Noether's definition of the stress tensor,\footnote{Our convention: $g^{00}=-1$, $g^{11}=g^{22}=g^{33}=1$.}
\begin{equation}\label{emt}
T^{\kappa\nu} = \frac{\partial \mathcal{L}}{\partial(\partial_\kappa F)}\partial^\nu F - g^{\kappa\nu}\mathcal{L},
\end{equation}
where $F$ generically denotes the variables in the Lagrangian. Equating the momentum density $\rho v_i$ to $T_{0i}$ immediately fixes the dependence of velocity on $\theta$ as
\begin{equation}\label{vdt}
v_i = -\partial_i \theta.
\end{equation}
However, for such a $\vec{v}$, vorticity vanishes: $\vec{\omega} = \vec{\nabla}\times \vec{v}=0$. We can overcome this problem by extending \eqref{vdt} to
\begin{equation}\label{vdtab}
v_i = -\partial_i \theta + \alpha \partial_i \beta,
\end{equation}
which is precisely the Clebsch decomposition of a vector field in terms of three scalar fields. In order to ensure $T_{0i}=\rho v_i$, one can easily check that the Lagrangian should be modified to
\begin{equation} \label{lang-fd01}
\mathcal{L} =  \rho (\partial_0 \theta - \alpha\partial_0\beta) - \Big( \frac{1}{2} \rho v^2(\theta,\alpha,\beta) + \rho \epsilon(\rho) \Big).
\end{equation}
The same argument, $T_{0i}=\rho v_i$, further generalises \eqref{vdtab} and \eqref{lang-fd01} for the case of nonisentropic fluids to
\begin{gather}
\label{v-fd}
v_i = -\partial_i \theta + \alpha\partial_i \beta + \lambda\partial_i s,\\
\label{lang-fd}
\mathcal{L} =  \rho (\partial_0 \theta - \alpha\partial_0\beta - \lambda\partial_0 s) - \Big( \frac{1}{2} \rho v^2(\theta,\alpha,\beta,\lambda,s) + \rho \epsilon(\rho,s) \Big),
\end{gather}
where $\rho$, $s$, $\theta$, $\lambda$, $\alpha$ and $\beta$ are the basic fields in the Lagrangian and the intensive variables, pressure $p$ and temperature $T$, of the fluid are obtained from $\epsilon(\rho, s)$:
\begin{equation} \label{eqpt}
p = \rho^2 \frac{\partial \epsilon}{\partial \rho}, \quad T = \frac{\partial \epsilon}{\partial s}.
\end{equation}

Extending this approach further to ideal MHD, we postulate the action as
\begin{equation}\label{actmhd}\begin{split}
&{} S[\rho,s,\theta,\lambda,\alpha,\beta,K_i,B_i] = \int\!\ud t\, \ud^3 x\, \mathcal{L}\\ &= \int\!\ud t\, \ud^3 x \Big( - \theta \partial_0 \rho - \lambda \rho \partial_0 s - \alpha \rho \partial_0 \beta - K_i \partial_0 B_i \\ &{}\qquad\qquad\quad - \frac{1}{2} \rho v^2(\rho,s,\theta,\lambda,\alpha,\beta,K_i,B_i) - \rho \epsilon(\rho,s) -\frac{B^2}{2\mu} \Big),\end{split}
\end{equation}
with the Clebsch-type decomposition for the velocity as
\begin{equation}
v_i = -\partial_i \theta + \lambda\partial_i s + \alpha\partial_i \beta + \frac{1}{\rho} f_i(K_j,B_j,\partial_m K_j, \partial_m B_j).
\end{equation}
The new entry here is that of the magnetic field $\vec{B}$ and another field $\vec{K}$, while $\vec{f}$ is some function of $\vec{K}$, $\vec{B}$ and their derivatives, as indicated, to be chosen appropriately so as to satisfy MHD equations \eqref{eqrho}--\eqref{eqv}.

Now we find the Euler-Lagrange equations following from the action \eqref{actmhd}. Variations with respect to the fields $\theta$ and $\lambda$ reproduce \eqref{eqrho} and \eqref{eqs}, respectively, while the variation with respect to $\alpha$ gives
\begin{equation}
\label{eqbeta} \partial_0 \beta + v_i \partial_i \beta = 0.
\end{equation}
Variations of the action with respect to $\beta$ and $s$, along with the use of \eqref{eqrho}, yield
\begin{gather}
\label{eqalpha} \partial_0 \alpha + v_i \partial_i \alpha = 0,\\
\label{eqlambda} \partial_0 \lambda + v_i \partial_i \lambda - \frac{\partial \epsilon}{\partial s} = 0.
\end{gather}
Similarly, varying $\rho$ and utilising \eqref{eqs} and \eqref{eqbeta} give
\begin{gather}
\label{eqtheta} \partial_0 \theta + v_i \partial_i \theta + \frac{v^2}{2} - \epsilon - \rho \frac{\partial \epsilon}{\partial \rho} = 0.
\end{gather}
Finally, we take variations with respect to $K_i$ and $B_i$, then we get
\begin{gather}
\label{eqB1} 
\partial_0 B_i + v_j \frac{\partial f_j}{\partial K_i} - \partial_m\left( v_j\frac{\partial f_j}{\partial(\partial_m K_i)} \right) = 0, \\
\label{eqK1} 
\partial_0 K_i -\frac{B_i}{\mu} - v_j \frac{\partial f_j}{\partial B_i} + \partial_m\left( v_j\frac{\partial f_j}{\partial(\partial_m B_i)} \right) = 0.
\end{gather}
The form of $f_i(K_j,B_j,\partial_m K_j, \partial_m B_j)$ is now fixed by requiring that \eqref{eqB1} should reproduce \eqref{eqB}, which imposes conditions on $\vec{f}$:
\begin{gather}
\label{f1}
\partial_m\left( \frac{\partial f_j}{\partial(\partial_m K_i)} \right) - \frac{\partial f_j}{\partial K_i} = \delta_{ij} \partial_m B_m - \partial_j B_i, \\
\label{f2}
\frac{\partial f_j}{\partial(\partial_m K_i)} = \delta_{ij} B_m - \delta_{jm} B_i.
\end{gather}
Acting $\partial_m$ on \eqref{f2} and then subtracting \eqref{f1} gives
\begin{equation}
\label{f3}
\frac{\partial f_j}{\partial K_i} = 0,
\end{equation}
which restricts $\vec{f}$ to
\begin{equation}
f_i = c_1 B_i \partial_j K_j + c_2 B_j \partial_i K_j + c_3 B_j \partial_j K_i.
\end{equation}
Equation \eqref{f2} then immediately fixes the coefficients: $c_1 = 0$, $c_2 = -1$ and $c_3 = 1$. Thus,
\begin{equation}\label{f}
f_i =  B_j (\partial_j K_i - \partial_i K_j).
\end{equation}
This $\vec{f}$ reduces \eqref{eqK1} to
\begin{equation}
\label{eqK} \partial_0 K_i - (B_i /\mu) -v_j (\partial_i K_j - \partial_j K_i) = 0,
\end{equation}
and it fixes the Clebsch decomposition for the velocity as
\begin{equation}\label{eqv1}
v_i = -\partial_i \theta + \lambda\partial_i s + \alpha\partial_i \beta + \frac{1}{\rho} B_j ( \partial_j K_i - \partial_i K_j).
\end{equation}
Similar decompositions have earlier appeared in \cite{Morrison1982} for various kinds of fluids.\footnote{
Clebsch decompositions of velocity for ideal MHD given in \cite{Morrison1982} is $\rho v_i = (\partial_i T_j)B_j - B_j \partial_j T_i - T_i \partial_j B_j + \rho \partial_i \phi + \sigma \partial_i\psi$. Writing $\sigma/\rho=s$, this can be rewritten as
$v_i = \partial_i \phi + s\partial_i \psi + \frac{1}{\rho} B_j (\partial_i T_j - \partial_j T_i) - T_i (\nabla \cdot \vec{B})$. To make comparison with our velocity decomposition, let us write $\psi$ as $-\lambda$, $\phi$ as $-\theta +s\lambda$ and $T_i$ as $K_i$. Then it looks $v_i = -\partial_i \theta + \lambda \partial_i s + \frac{1}{\rho} B_j (\partial_i K_j - \partial_j K_i) - K_i (\nabla \cdot \vec{B})$, in which the last term can be dropped by enforcing Gauss's law. This velocity decomposition is then just the same as obtained by us in \eqref{eqv1}, apart from the $\alpha\partial_i \beta$ term in \eqref{eqv1} which is the imposition of Lin's constraint to incorporate vortical flows.
}

Now it is only the MHD Euler equation, Eq.~\eqref{eqv}, which remains to be derived. For that we act $(\partial_0 + \vec{v}\cdot\vec{\nabla})$ on Eq.~\eqref{eqv1}:
\begin{equation}
(\partial_0 + \vec{v}\cdot\vec{\nabla})v_i = (\partial_0 + \vec{v}\cdot\vec{\nabla})\Big[-\partial_i \theta + \lambda\partial_i s + \alpha\partial_i \beta + \frac{1}{\rho} B_j ( \partial_j K_i - \partial_i K_j)\Big].
\end{equation}
Using Eqs.~\eqref{eqrho}--\eqref{eqB}, \eqref{eqbeta}--\eqref{eqtheta} and \eqref{eqK} on the right-hand side to eliminate the time-derivatives reproduces, after some lengthy algebra,
\begin{equation}\label{x2}
(\partial_0 + \vec{v}\cdot \vec{\nabla}) v_i = -\frac{1}{\rho} \partial_i p + \frac{1}{\mu \rho} [(\vec{\nabla}\times \vec{B}) \times \vec{B}]_i + \frac{1}{\rho} (\vec{\nabla}\cdot \vec{B}) v_j(\partial_j K_i - \partial_i K_j),
\end{equation}
where the first term on the right-hand side is obtained using
\begin{equation}\label{x1}
-\partial_i \Big( \epsilon + \rho \frac{\partial \epsilon}{\partial\rho}\Big) + \frac{\partial\epsilon}{\partial s}\partial_i s = -\frac{1}{\rho} \partial_i p,
\end{equation}
which follows from the definition of pressure given in \eqref{eqpt}. At this stage, it is imperative to use \eqref{gaussB} to get rid of the last term on the right-hand side of \eqref{x2} and identify it with the Euler equation, Eq.~\eqref{eqv}. This completes the Lagrangian formulation of the our action \eqref{actmhd}.

It is possible to relate our action \eqref{actmhd} to the one considered in \cite{BO2000}. We consider the term involving velocity in the action \eqref{actmhd}:
\begin{equation}
\begin{aligned}
- \frac{1}{2} \rho v^2 &= \frac{1}{2} \rho v^2 - \rho v^2\\
 &= \frac{1}{2} \rho v^2 - \rho v_i \Big[ -\partial_i \theta + \lambda\partial_i s + \alpha\partial_i \beta + \frac{1}{\rho} B_j ( \partial_j K_i - \partial_i K_j) \Big],
\end{aligned}
\end{equation}
where we have made selective use of the Clebsch decomposition of velocity \eqref{eqv1}. Treating the remaining $v_i$ as independent fields, the action \eqref{actmhd} reads
\begin{equation}\label{actmhd1}\begin{split}
S &= \int\!\ud t\, \ud^3 x \Big( - \theta \partial_0 \rho - \lambda \rho \partial_0 s - \alpha \rho \partial_0 \beta - K_i \partial_0 B_i - \rho \epsilon(\rho,s) -\frac{B^2}{2\mu} \\ &{}\qquad\qquad + \frac{1}{2} \rho v^2 - \rho v_i \Big[ -\partial_i \theta + \lambda\partial_i s + \alpha\partial_i \beta + \frac{1}{\rho} B_j ( \partial_j K_i - \partial_i K_j) \Big] \Big),\end{split}
\end{equation}
which can be rearranged as
\begin{equation}\begin{split}
S &= \int\!\ud t\, \ud^3 x \Big( \frac{1}{2} \rho v^2 - \rho \epsilon(\rho,s) -\frac{B^2}{2\mu} - \theta \left[ \partial_0 \rho + \vec{\nabla} \cdot (\rho \vec{v}) \right] - \lambda \rho \left[ \partial_0 s + \vec{v} \cdot \vec{\nabla} s \right] \\ &{}\qquad\qquad\quad  - \alpha \rho \left[ \partial_0 \beta + \vec{v} \cdot \vec{\nabla}\beta \right] - \vec{K}\cdot \left[ \partial_0 \vec{B} - \vec{\nabla} \times (\vec{v} \times \vec{B}) \right] \Big).\end{split}
\end{equation}
This is the action considered in \cite{BO2000} in which $\vec{v}$ is also a basic field, whereas in our action \eqref{actmhd} $\vec{v}$ is not a basic field.

%%%%%%%%%%%%%%%%%%%%%%%%%%%%%%%%%%%%%%%%%%%%%%%%%%%%%%%%%%%%%%%%%%

\section{\label{mhdham}Hamiltonian formulation of MHD}

In this section we shall provide a new Hamiltonian formulation of ideal MHD based on Dirac's constraint analysis \cite{Dirac1964}. It will involve two complementary descriptions. First we shall work with the usual canonical Hamiltonian but the brackets have noncanonical form. The brackets are systematically obtained by the Dirac's algorithm. They are essentially Dirac brackets and reproduce the standard noncanonial brackets of MHD found in the literature \cite{MG1980}. In the second version we shall give a noncanonical Hamiltonian but all the brackets are canonical. We show here that it is basically a trade-off between a canonical Hamiltonian with noncanonical brackets and a noncanonical Hamiltonian with canonical brackets.

Now we proceed to carry out a Hamiltonian formulation for the action \eqref{actmhd}. The momenta conjugate to $\rho$, $s$, $\beta$, $B_i$, $\theta$, $\lambda$, $\alpha$ and $K_i$ are, respectively,
\begin{equation}\label{momenta}
\begin{gathered}
\pi_\rho = -\theta, \quad \pi_s = -\lambda\rho, \quad \pi_\beta = -\alpha\rho, \quad \pi^B_i = - K_i, \\
\pi_\theta = 0, \quad \pi_\lambda = 0, \quad \pi_\alpha = 0, \quad \pi^K_i = 0, 
\end{gathered}
\end{equation}
while the canonical Hamiltonian is
\begin{equation}\label{hammhd}
H = \int\!\ud^3 x\, \mathcal{H} = \int\!\ud^3 x \Big( \frac{1}{2} \rho v^2(\rho,s,\theta,\lambda,\alpha,\beta,K_i,B_i) + \rho \epsilon(\rho,s) + \frac{B^2}{2\mu} \Big).
\end{equation}
Since this is a constrained system we will follow Dirac's algorithm \cite{Dirac1964} to construct a Hamiltonian formulation of MHD. The primary constraints of the theory follow from \eqref{momenta}, which we label as
\begin{equation}\label{cons}
\begin{gathered}
\Omega_1 = \pi_\rho + \theta, \quad \Omega_2 = \pi_s + \lambda\rho, \quad \Omega_3 = \pi_\beta + \alpha\rho, \\
\Omega_{3+i} = \pi^B_i + K_i,\; i=1,2,3,\\
\Omega_7 = \pi_\theta, \quad \Omega_8 = \pi_\lambda, \quad \Omega_9 = \pi_\alpha, \\
\Omega_{9+i} = \pi^K_i,\; i=1,2,3.
\end{gathered}
\end{equation}
All these 12 constraints are second-class as seen from their Poisson brackets. We now construct the constraint matrix of the Poisson brackets,\footnote{All brackets are equal-time, so time argument is omitted for convenience.}
\begin{equation}
\Lambda_{a,b}(\vec{x},\vec{x}') = \{\Omega_a(\vec{x}), \Omega_b(\vec{x}')\},\quad a,b=1,\ldots,12,
\end{equation}
which has the following nonvanishing components:
\begin{equation}
\begin{gathered}
\Lambda_{1,2}(\vec{x},\vec{x}') = -\lambda \delta(\vec{x}-\vec{x}'), \quad
\Lambda_{1,3}(\vec{x},\vec{x}') = -\alpha \delta(\vec{x}-\vec{x}'), \\
\Lambda_{1,7}(\vec{x},\vec{x}') =  \delta(\vec{x}-\vec{x}'), \quad
\Lambda_{2,8}(\vec{x},\vec{x}') = \Lambda_{3,9}(\vec{x},\vec{x}') = \rho \delta(\vec{x}-\vec{x}'), \\
\Lambda_{3+i,9+j}(\vec{x},\vec{x}') = \delta_{ij} \delta(\vec{x}-\vec{x}'),\quad i,j=1,2,3.
\end{gathered}
\end{equation}
The inverse matrix, $\Lambda^{-1}(\vec{x},\vec{x}')$, defined through
\begin{equation}
\int\!\ud y^3 \Lambda^{-1}_{a,b}(\vec{x},\vec{y}) \Lambda_{b,c}(\vec{y},\vec{x}') = \delta_{ac} \delta(\vec{x}-\vec{x}'),
\end{equation}
has the nonvanishing components:
\begin{equation}
\begin{gathered}
\Lambda^{-1}_{1,7}(\vec{x},\vec{x}') = -\delta(\vec{x}-\vec{x}'), \quad
\Lambda^{-1}_{2,8}(\vec{x},\vec{x}') = \Lambda^{-1}_{3,9}(\vec{x},\vec{x}') = -\frac{1}{\rho} \delta(\vec{x}-\vec{x}'), \\
\Lambda^{-1}_{3+i,9+j}(\vec{x},\vec{x}') = - \delta_{ij}\delta(\vec{x}-\vec{x}'), \quad i,j=1,2,3, \\
\Lambda^{-1}_{7,8}(\vec{x},\vec{x}') = -\frac{\lambda}{\rho} \delta(\vec{x}-\vec{x}'),\quad
\Lambda^{-1}_{7,9}(\vec{x},\vec{x}') = -\frac{\alpha}{\rho} \delta(\vec{x}-\vec{x}').
\end{gathered}
\end{equation}
In Dirac's procedure, the canonical Hamiltonian \eqref{hammhd} is replaced by the total Hamiltonian
\begin{equation}\label{hamtmhd}
\begin{split}
H_\mathrm{T} &= \int\!\ud^3 x \Big( \mathcal{H} + C_a \Omega_a\Big)\\
&= \int\!\ud^3 x \Big( \frac{1}{2} \rho v^2(\rho,s,\theta,\lambda,\alpha,\beta,K_i,B_i) + \rho \epsilon(\rho,s) + \frac{B^2}{2\mu} + C_a \Omega_a\Big),
\end{split}
\end{equation}
where $C_a$, $a=1,\ldots, 12$, are the Lagrange multiplier fields implementing the constraints \eqref{cons}. Since the constraint matrix $\Lambda$ is invertible, it is a second-class system. Now there are two possibilities. The constraints may be eliminated by working with Dirac brackets instead of Poisson brackets. This will give a formulation where the Hamiltonian retains its canonical structure \eqref{hammhd} but the basic algebra is given by the Dirac brackets. The other option is to fix the multipliers in \eqref{hamtmhd} by requiring time-conservation of the constraints. Then we have a formulation involving the total (noncanonical) Hamiltonian \eqref{hamtmhd} but all brackets are canonical. The second option is discussed later.

\subsection{Hamiltonian formulation in terms of noncanonical brackets}

In the first option, the second-class constraints (\ref{cons}) can be eliminated by computing the Dirac brackets, denoted by a star, which are defined in terms of the usual canonical (Poisson) brackets as
\begin{equation}\label{new}
\begin{split}
\{F(\vec{x}), G(\vec{x}')\}^* &= \{F(\vec{x}), G(\vec{x}')\}\\
&{}\quad - \int\!\ud^3 y\, \ud^3 z\, \{F(\vec{x}), \Omega_a(\vec{y})\} \Lambda^{-1}_{a,b}(\vec{y},\vec{z}) \{\Omega_b(\vec{z}), G(\vec{x}')\}.
\end{split}
\end{equation}
In our case, the nonvanishing Dirac brackets among various fields turn out to be
\begin{equation}\label{brac}
\begin{gathered}
\{\rho(\vec{x}), \theta(\vec{x}')\}^* = - \delta(\vec{x}-\vec{x}'), \quad
\{\lambda(\vec{x}), \theta(\vec{x}')\}^* = \frac{\lambda}{\rho} \delta(\vec{x}-\vec{x}'), \\
\{\alpha(\vec{x}), \theta(\vec{x}')\}^* = \frac{\alpha}{\rho} \delta(\vec{x}-\vec{x}'), \\
\{\lambda(\vec{x}), s(\vec{x}')\}^* = \{\alpha(\vec{x}), \beta(\vec{x}')\}^* = \frac{1}{\rho} \delta(\vec{x}-\vec{x}'), \\
\{K_i(\vec{x}), B_j(\vec{x}')\}^* = \delta_{ij}\delta(\vec{x}-\vec{x}').
\end{gathered}
\end{equation}
The physical fields $\rho$, $s$ and $B_i$ have vanishing brackets among themselves.

At this stage, we can make a consistency check. From the brackets \eqref{brac} one can easily identify the canonical pairs, which are $(\theta,\rho)$, $(\rho\lambda,s)$, $(\rho\alpha, \beta)$ and $(K_i,B_i)$.
 The same set of pairs can also be identified from the action \eqref{actmhd} itself.

From the construction (\ref{new}) it is seen that the constraints (\ref{cons}) have vanishing Dirac brackets with all variables appearing in the total Hamiltonian (\ref{hamtmhd}). Effectively, therefore, the constraints may be strongly eliminated from the phase space. The total Hamiltonian then reduces to the canonical form (\ref{hammhd}). It is now advantageous to obtain the Dirac brackets among $\rho$, $s$, $v_i$ and $B_i$, because then we can use the Hamiltonian \eqref{hammhd} explicitly expressed in terms of these physical fields,
\begin{equation}\label{hammhd2}
H = \int\!\ud^3 x \Big( \frac{1}{2} \rho v^2 + \rho \epsilon(\rho,s) + \frac{B^2}{2\mu} \Big),
\end{equation}
to obtain the equations for MHD. As mentioned earlier, $\rho$, $s$ and $B_i$ have vanishing brackets among themselves. So we need to find the brackets of $v_i$ with these fields. Using brackets \eqref{brac} and Eq.~\eqref{eqv1} it is straightforward to see that
\begin{gather}
\label{bravrho} \{v_i(\vec{x}), \rho(\vec{x}')\}^* = -\partial_i \delta(\vec{x}-\vec{x}'), \\
\label{bravs} \{v_i(\vec{x}), s(\vec{x}')\}^* = \frac{\partial_i s}{\rho} \delta(\vec{x}-\vec{x}'), \\
\label{braBv} \{B_i(\vec{x}), v_j(\vec{x}')\}^* = \delta_{ij} \left( \frac{B_k}{\rho} \right)_{x'} \partial_k \delta(\vec{x}-\vec{x}') - \left( \frac{B_i}{\rho} \right)_{x'} \partial_j \delta(\vec{x}-\vec{x}').
\end{gather}
The $v_i$--$v_j$ bracket is somewhat involved, so we give a few intermediate steps. Use of brackets \eqref{brac} and Eq.~\eqref{eqv1} also yields
\begin{gather}
\label{bravtheta} \{v_i(\vec{x}), \theta(\vec{x}')\}^* = \frac{1}{\rho}\left( \lambda \partial_i s + \alpha \partial_i \beta -\frac{1}{\rho} B_j [\partial_i K_j - \partial_j K_i]\right) \delta(\vec{x}-\vec{x}'), \\
\label{bravlambda} \{v_i(\vec{x}), \lambda(\vec{x}')\}^* = \frac{1}{\rho} \partial_i \lambda\, \delta(\vec{x}-\vec{x}'), \\
\label{bravalpha} \{v_i(\vec{x}), \alpha(\vec{x}')\}^* = \frac{1}{\rho} \partial_i \alpha\, \delta(\vec{x}-\vec{x}'), \\
\label{bravbeta} \{v_i(\vec{x}), \beta(\vec{x}')\}^* = \frac{1}{\rho}\partial_i  \beta\, \delta(\vec{x}-\vec{x}'), \\
\label{bravK} \{v_i(\vec{x}), K_j(\vec{x}')\}^* = \frac{1}{\rho} (\partial_i K_j - \partial_j K_i) \delta(\vec{x}-\vec{x}').
\end{gather}
Equation \eqref{eqv1} also gives the following expression for vorticity in terms of basic fields:
\begin{equation}\label{vort}
\begin{split}
\omega_{ij} &\equiv \partial_i v_j - \partial_j v_i \\
&= \partial_i \lambda \partial_j s + \partial_i \alpha \partial_j \beta + \partial_i(B_m/\rho)[\partial_m K_j -\partial_j K_m] + (B_m/\rho)\partial_i \partial_m K_j -\langle i\leftrightarrow j\rangle,
\end{split}
\end{equation}
where $\langle i\leftrightarrow j\rangle$ stands for the previous terms with $i$ and $j$ interchanged. Now we proceed to evaluate the $v_i$--$v_j$ bracket:
\begin{equation}
\{v_i(\vec{x}), v_j(\vec{x}')\}^* = \left\{v_i(\vec{x}), \Big[-\partial_j \theta + \lambda\partial_j s + \alpha\partial_j \beta + \frac{1}{\rho} B_m ( \partial_m K_j - \partial_j K_m)\Big](\vec{x}')\right\}^*.
\end{equation}
Brackets \eqref{bravrho}--\eqref{bravK} are used to simplify the right-hand side of the above equation and finally we use \eqref{vort}. Then it gives
\begin{gather}
\label{bravv1} \{v_i(\vec{x}), v_j(\vec{x}')\}^* = \frac{1}{\rho} [\omega_{ij} - (\vec{\nabla}\cdot\vec{B}) (\partial_i K_j - \partial_j K_i)/\rho] \delta(\vec{x}-\vec{x}').
\end{gather}
Thus we see that unless we impose Gauss's law \eqref{gaussB}, we cannot express $v_i$--$v_j$ bracket solely in terms of physical variables. Imposing \eqref{gaussB}, $\vec{\nabla}\cdot\vec{B}=0$, we then have
\begin{equation}
\label{bravv} \{v_i(\vec{x}), v_j(\vec{x}')\}^* = \frac{\omega_{ij}}{\rho} \delta(\vec{x}-\vec{x}').
\end{equation}
It should be stated that imposition of $\vec{\nabla}\cdot\vec{B}=0$ is consistent with the algebra \eqref{braBv} as may easily be checked by taking a divergence on both sides of that equation. Thus the nonvanishing brackets among the physical variables $\rho$, $s$, $B_i$ and $v_i$ are \eqref{bravrho}--\eqref{braBv} and \eqref{bravv}. These nonvanishing (Dirac) brackets are just the noncanonical brackets of MHD posited in \cite{MG1980}.

From these brackets and the Hamiltonian \eqref{hammhd2}, the MHD equations \eqref{eqrho}--\eqref{eqv} follow in the usual way:
\begin{equation}
\begin{gathered}
\partial_0 \rho = \{\rho, H\}^*, \quad
\partial_0 s = \{s, H\}^*, \\
\partial_0 B_i = \{B_i, H\}^*, \quad
\partial_0 v_i = \{v_i, H\}^*.
\end{gathered}
\end{equation}
For example,
\begin{equation}
\partial_0 v_i (t,\vec{x}) = \left\{ v_i (t,\vec{x}), H\right\}^* = \left\{ v_i (t,\vec{x}), \int\!\ud^3 x' \Big( \frac{1}{2} \rho v^2 + \rho \epsilon(\rho,s) + \frac{B^2}{2\mu} \Big)(t,\vec{x}')\right\}^*
\end{equation}
yields $\partial_0 v_i =-v_j\partial_j v_i -(1/\rho)\partial_i p + (1/\mu\rho)(B_j \partial_j B_i - B_j\partial_i B_j)$, which is the Euler equation \eqref{eqv} in component form.

It is worthwhile to mention here the role of Gauss's law, Eq.~\eqref{gaussB}. It is necessary to impose this condition ($\vec{\nabla}\cdot \vec{B} =0$) to obtain the Euler equation \eqref{eqv} from the action \eqref{actmhd}. This condition is also necessary to express the $v_i$--$v_j$ bracket solely in terms of physical fields. It is interesting to note that this law is not respected by the $K$--$B$ bracket given in \eqref{brac}, as one can easily see: $\{K_i(\vec{x}), (\vec{\nabla}\cdot \vec{B})(\vec{x}')\}^* = -\partial_i \delta(\vec{x}-\vec{x}') \ne 0$. However, this is not a problem as $\vec{K}$ is not a physical field. The physical fields are $\rho$, $s$, $v_i$ and $B_i$, which satisfy the brackets \eqref{bravrho}--\eqref{braBv} and \eqref{bravv}. As already stated, from \eqref{braBv}, which is the only nonvanishing bracket involving $\vec{B}$, with a fluid variable, it follows that $\{\partial_i B_i(\vec{x}), v_j(\vec{x}')\}^* = 0$. Thus, the brackets among the physical fields, \eqref{bravrho}--\eqref{braBv} and \eqref{bravv}, indeed respect Gauss's law.

We mentioned below \eqref{brac} that $(\theta,\rho)$, $(\rho\lambda,s)$, $(\rho\alpha, \beta)$ and $(K_i,B_i)$ are the canonical pairs. So one can define
\begin{equation}
\lambda' = \rho\lambda, \quad \alpha' = \rho\alpha,
\end{equation}
and use these $\lambda'$ and $\alpha'$ instead of $\lambda$ and $\alpha$. Then the Hamiltonian and the velocity decomposition look like
\begin{gather}
\label{n1} H = \int\!\ud^3 x \Big( \frac{1}{2} \rho v^2(\rho,s,\theta,\lambda',\alpha',\beta,K_i,B_i) + \rho \epsilon(\rho,s) + \frac{B^2}{2\mu} \Big), \\
\label{n2} v_i = -\partial_i \theta + \frac{\lambda'}{\rho}\partial_i s + \frac{\alpha'}{\rho}\partial_i \beta + \frac{1}{\rho} B_j ( \partial_j K_i - \partial_i K_j).
\end{gather}
The MHD equations then follow from \eqref{n1} and \eqref{n2} using the following nonvanishing brackets:
\begin{equation}
\begin{gathered}
\{\theta(\vec{x}), \rho(\vec{x}')\}^* = \{\lambda'(\vec{x}), s(\vec{x}')\}^*= \{\alpha'(\vec{x}), \beta(\vec{x}')\}^* = \delta(\vec{x}-\vec{x}'), \\
\{K_i(\vec{x}), B_j(\vec{x}')\}^* = \delta_{ij}\delta(\vec{x}-\vec{x}').
\end{gathered}
\end{equation}
Similar treatment has earlier appeared in \cite{Henyey1982}, though the results were obtained from the different point of view.

A canonical analysis where the brackets involving the basic MHD variables are canonical is possible which would be the goal of the next subsection.

\subsection{Hamiltonian formulation in terms of canonical brackets}

Now we discuss the second option which will involve the total (noncanonical) Hamiltonian but all the brackets will be canonical. For that we have to fix the multipliers $C_a$ appearing in \eqref{hamtmhd}. Conserving all the primary constraints with time,
\begin{equation}
\partial_0 \Omega_a = \left\{ \Omega_a, H_\mathrm{T}\right\} = 0,
\end{equation}
gives conditions on $C_a$. For example, $\left\{ \Omega_1, H_\mathrm{T}\right\} = 0$ gives
\begin{equation}
C_7 -\lambda C_2 -\alpha C_3 -\frac{v^2}{2} + \frac{1}{\rho} v_i B_j (\partial_j K_i - \partial_i K_j) -\epsilon -\rho \frac{\partial \epsilon}{\partial \rho} = 0.
\end{equation}
We get 12 such conditions in total corresponding to the 12 constraints $\Omega_a$. These conditions uniquely fix the multipliers:
\begin{equation}\label{Cs}
\begin{gathered}
C_1 = -\partial_i (\rho v_i), \quad
C_2 =  -v_i \partial_i s, \quad
C_3 = -v_i \partial_i \beta, \\
C_{3+i} = \partial_j(v_i B_j - v_j B_i), \; i=1,2,3,\\
C_7 = -\frac{v^2}{2} - v_i \partial_i \theta + \epsilon + \rho \frac{\partial \epsilon}{\partial \rho}, \quad
C_8 = -v_i \partial_i \lambda + \frac{\partial \epsilon}{\partial s}, \quad
C_9 = -v_i \partial_i \alpha, \\
C_{9+i} = (B_i /\mu) + v_j (\partial_i K_j - \partial_j K_i), \; i=1,2,3,
\end{gathered}
\end{equation}
where $v_i$ appearing on the right-hand sides of these equations is expressed in terms of other fields as given in \eqref{eqv1}. Equations of motion now follow using the standard Poisson brackets ($\{ \rho(\vec{x}),\pi_\rho (\vec{x}')\} = \delta(\vec{x}-\vec{x}')$, etc.)\ and the Hamiltonian $H_\mathrm{T}$ given in \eqref{hamtmhd} with the multipliers $C_a$ as given \eqref{Cs} and $v_i$ as given in \eqref{eqv1}. Obtaining Eqs.~\eqref{eqrho}--\eqref{eqB} is just straightforward. Derivation of Euler equation is somewhat involved, which we now explicitly demonstrate. Using $v_i$ from \eqref{eqv1}, we have
\begin{equation}
\begin{split}
\partial_0 v_i &= \left\{ v_i, H_\mathrm{T} \right\} \\
&= \left\{ \Big(-\partial_i \theta + \lambda\partial_i s + \alpha\partial_i \beta + \frac{1}{\rho} B_j (\partial_j K_i - \partial_i K_j)\Big), H_\mathrm{T}\right\}.
\end{split}
\end{equation}
Noting that it is only the $C_a \Omega_a$ term on the right-hand side of \eqref{hamtmhd} which involves momenta, it then follows (using standard Poisson brackets) that
\begin{equation}
\begin{split}
\partial_0 v_i &= -\partial_i C_7 + \lambda\partial_i C_2 + (\partial_i s)C_8 + \alpha \partial_i C_3 + (\partial_i \beta) C_9 \\
&{} \qquad + \frac{1}{\rho} (\partial_j K_i -\partial_i K_j) \left(C_{3+j} - B_j C_1/\rho \right) + \frac{1}{\rho} B_j (\partial_j C_{9+i} - \partial_i C_{9+j}).
\end{split}
\end{equation}
Now we make use of \eqref{Cs}, use \eqref{x1} and finally use \eqref{eqv1} to express the right-hand side in terms of velocity. Then we get
\begin{equation}
\partial_0 v_i = - v_j\partial_j v_i -\frac{1}{\rho} \partial_i p + \frac{1}{\mu \rho} [(\vec{\nabla}\times \vec{B}) \times \vec{B}]_i + \frac{1}{\rho} (\vec{\nabla}\cdot \vec{B}) v_j(\partial_j K_i - \partial_i K_j),
\end{equation}
which is just Eq.~\eqref{x2} obtained earlier. Once we use Gauss's law, it is just the Euler equation \eqref{eqv}.

Thus, we have shown that MHD equations can be obtained from a canonical Hamiltonian \eqref{hammhd} using the noncanonical brackets, \eqref{bravrho}--\eqref{braBv} and \eqref{bravv}, or from a noncanonical Hamiltonian \eqref{hamtmhd} using the canonical brackets.

%%%%%%%%%%%%%%%%%%%%%%%%%%%%%%%%%%%%%%%%%%%%%%%%%%%%%%%%%%%%%%%%%%

\section{\label{mhdinc}Hamiltonian formulation of incompressible MHD}

In this section we discuss the case of incompressible MHD, i.e.\ the fluid density is constant (in time) and uniform (in space): $\rho = \rho_0$. We start with the MHD action \eqref{actmhd} and incorporate the incompressibility constraint by a multiplier:
\begin{equation}\label{actmhdinc}\begin{split}
S^\mathrm{inc} &= \int\!\ud t\, \ud^3 x \Big( - \theta \partial_0 \rho - \lambda \rho \partial_0 s - \alpha \rho \partial_0 \beta - K_i \partial_0 B_i \\ &{}\qquad\qquad\quad - \frac{1}{2} \rho v^2 - \rho \epsilon(\rho,s) -\frac{B^2}{2\mu} + F(\rho-\rho_0) \Big),\end{split}
\end{equation}
Since $F$ is a multiplier field, the corresponding conjugate momentum $\pi_F$ is zero and we have, in addition to the constraints $\Omega_a$, $a=1,\ldots, 12$, given in \eqref{cons}, another primary constraint:
\begin{equation}\label{con0}
\tilde{\Omega}_{0} = \pi_F \approx 0.
\end{equation}
The canonical Hamiltonian now is
\begin{equation}\label{hammhdinc}
H^\mathrm{inc} = \int\!\ud^3 x\, \mathcal{H}^\mathrm{inc} = \int\!\ud^3 x \Big( \frac{1}{2} \rho v^2 + \rho \epsilon(\rho,s) + \frac{B^2}{2\mu} - F(\rho-\rho_0) \Big).
\end{equation}
In the Dirac's procedure, the canonical Hamiltonian \eqref{hammhdinc} is replaced by the total Hamiltonian
\begin{equation}\label{hamtmhdinc}
H^\mathrm{inc}_\mathrm{T} = \int\!\ud^3 x \Big( \frac{1}{2} \rho v^2 + \rho \epsilon(\rho,s) + \frac{B^2}{2\mu} - F(\rho-\rho_0) + C_a \Omega_a + \tilde{C}_{0}\tilde{\Omega}_{0} \Big),
\end{equation}
where $C_a$, $a=1,\ldots, 12$, and  $\tilde{C}_{0}$ are the Lagrange multiplier fields corresponding to the constraints \eqref{cons} and \eqref{con0}. Conserving the constraint $\tilde{\Omega}_{0}$ with time yields the incompressibility condition:
\begin{equation}\label{inc}
\tilde{\Omega}_{1} \equiv \rho-\rho_0 \approx 0,
\end{equation}
and conserving $\tilde{\Omega}_{1}$ with time fixes the Lagrange multiplier $C_1$:
\begin{equation}\label{c1}
C_1 = 0.
\end{equation}
Conservation of $\Omega_{7}$ with time yields $\partial_i (\rho v_i) + C_1 = 0$, which in view of \eqref{inc} and \eqref{c1} gives us another (secondary) constraint,
\begin{equation}\label{inc2}
\tilde{\Omega}_{2} \equiv \partial_i v_i \approx 0.
\end{equation}
Similarly, conserving $\tilde{\Omega}_{2}$ with time and making use of \eqref{c1} we get
\begin{equation}\label{new2}
\begin{aligned}
&\partial_i \Big[ -\partial_i C_7 + C_8 \partial_i s + \lambda \partial_i C_2 + C_9 \partial_i \beta + \alpha \partial_i C_3 \\ 
&{}\qquad + \frac{1}{\rho}(\partial_j K_i -\partial_i K_j)C_{3+j} + \frac{1}{\rho} B_j(\partial_j C_{9+i} - \partial_i C_{9+j}) \Big] \approx 0,
\end{aligned}
\end{equation}
which is a relation among various multipliers. This is now evident that the constraints $\Omega_a$, $a=1,\ldots, 12$, 
$\tilde{\Omega}_{1}$ and $\tilde{\Omega}_{2}$ form a set of 14 second-class constraints while the constraint $\tilde{\Omega}_{0}$ is first-class. Since $F$ is multiplier, we can discard the conjugate pair $(F, \pi_F)$ from the phase space altogether. Then the total Hamiltonian \eqref{hamtmhdinc} reduces to
\begin{equation}\label{hamtmhdinc2}
H^\mathrm{inc}_\mathrm{T} = \int\!\ud^3 x \Big( \frac{1}{2} \rho v^2 + \rho \epsilon(\rho,s) + \frac{B^2}{2\mu} + C_a \Omega_a \Big).
\end{equation}
Conservation of the remaining constraints yields conditions on the Lagrange multipliers which can be simplified to the following:
\begin{equation}\label{Cs2}
\begin{gathered}
C_2 =  -v_i \partial_i s, \quad
C_3 = -v_i \partial_i \beta, \quad
C_{3+i} = B_j \partial_j v_i - v_j \partial_j B_i, \; i=1,2,3,\\
C_7 = -\frac{v^2}{2} - v_i \partial_i \theta + \epsilon + \rho \frac{\partial \epsilon}{\partial \rho}, \quad
C_8 = -v_i \partial_i \lambda + \frac{\partial \epsilon}{\partial s}, \quad
C_9 = -v_i \partial_i \alpha, \\
C_{9+i} = (B_i /\mu) + v_j (\partial_i K_j - \partial_j K_i), \; i=1,2,3,
\end{gathered}
\end{equation}
where $v_i$ appearing on the right-hand sides of these equations is expressed in terms of other fields as given in \eqref{eqv1}. Since now all the multipliers have been fixed, various equations of motion follow from the total Hamiltonian \eqref{hamtmhdinc2} using the standard Poisson brackets. The equation for fluid density is $\partial_0 \rho = 0$, which is compatible with $\rho=\rho_0$, while others  are
\begin{gather}
\label{eqsinc} \partial_0 s + \vec{v} \cdot \vec{\nabla} s = 0, \\
\label{eqBinc} \partial_0 B_i = B_j \partial_j v_i - v_j \partial_j B_i, \\
\label{eqvinc} \partial_0 v_i = - v_j \partial_j v_i + \frac{1}{\mu \rho} B_j (\partial_j B_i - \partial_i B_j) - \rho \partial_i \left( \frac{\partial \epsilon}{\partial \rho} \right),
\end{gather}
where the last equation has been obtained using the Clebsch decomposition \eqref{eqv1}. These equations are basically the equations \eqref{eqrho}--\eqref{eqv} but subject to conditions \eqref{inc} and \eqref{inc2}. Also, since the multipliers have been fixed, Eq.~\eqref{new2} reduces to
\begin{equation}\label{new3}
\partial_i \left[ - v_j \partial_j v_i + \frac{1}{\mu \rho} B_j (\partial_j B_i - \partial_i B_j) - \rho \partial_i \left( \frac{\partial \epsilon}{\partial \rho} \right) \right] \approx 0,
\end{equation}
which is what also follows from \eqref{eqvinc} by acting $\partial_i$ on both sides and then using \eqref{inc2}.

In an alternative description of incompressible MHD we can compute the Dirac brackets and eliminate the second-class constraints $\Omega_a$, $a=1,\ldots, 12$, $\tilde{\Omega}_{1}$ and $\tilde{\Omega}_{2}$. However, we shall do the computation of Dirac brackets in two stages since such a computation for the general (compressible) case has already been done. Therefore we split the 14 second-class constraints into two sets, the first set contains  $\Omega_a$, $a=1,\ldots, 12$, while the second one contains $\tilde{\Omega}_{1}$ and $\tilde{\Omega}_{2}$. In the first step we compute the Dirac brackets with respect to the first set. Then the nonvanishing brackets among the physical variables $\rho$, $s$, $B_i$ and $v_i$ are \eqref{bravrho}--\eqref{braBv} and \eqref{bravv}, as computed in Sec.~\ref{mhdham}. Now to incorporate the second set we compute the brackets among the constraints
\begin{equation}
\tilde{\Lambda}_{\tilde{a},\tilde{b}}(\vec{x},\vec{x}') = \{\tilde{\Omega}_{\tilde{a}}(\vec{x}), \tilde{\Omega}_{\tilde{b}}(\vec{x}')\}^*,\quad \tilde{a},\tilde{b}=1,2,
\end{equation}
where the brackets on the right-hand side are the Dirac brackets computed with respect to the first set, i.e.\ \eqref{bravrho}--\eqref{braBv} and \eqref{bravv}. The nonvanishing elements of the constraint matrix thus are
\begin{equation}
\begin{gathered}
\tilde{\Lambda}_{1,2}(\vec{x},\vec{x}') = \nabla^2 \delta (\vec{x}-\vec{x}'), \quad
\tilde{\Lambda}_{2,2}(\vec{x},\vec{x}') = - \partial_i \left[ \frac{1}{\rho} \omega_{ij}\partial_j \delta (\vec{x}-\vec{x}')\right],
\end{gathered}
\end{equation}
while the nonvanishing components of the inverse matrix are
\begin{equation}
\begin{gathered}
\tilde{\Lambda}^{-1}_{1,1}(\vec{x},\vec{x}') = \Delta^{-1} \partial_i \left(\frac{1}{\rho} \omega_{ji}\partial_j \Delta^{-1} \delta (\vec{x}-\vec{x}')\right), \quad
\tilde{\Lambda}^{-1}_{1,2}(\vec{x},\vec{x}') = - \Delta^{-1} \delta (\vec{x}-\vec{x}'),
\end{gathered}
\end{equation}
where $\Delta^{-1}$ is the inverse of $\nabla^2$: if $\nabla^2 f(\vec{x}) = g(\vec{x})$, then $f(\vec{x}) = \Delta^{-1} g(\vec{x})$. That is, $\Delta^{-1} f(\vec{x}) = -(1/4\pi) \int \ud^3 x' f(\vec{x}')/|\vec{x}- \vec{x}'|$.
Now the Dirac brackets for incompressible MHD are computed as
\begin{equation}
\begin{split}
\{F(\vec{x}), G(\vec{x}')\}^*_\mathrm{inc} &= \{F(\vec{x}), G(\vec{x}')\}^*\\
&{}\quad - \int\!\ud^3 y\, \ud^3 z\, \{F(\vec{x}), \tilde{\Omega}_{\tilde{a}}(\vec{y})\}^* \tilde{\Lambda}^{-1}_{\tilde{a},\tilde{b}}(\vec{y},\vec{z}) \{\tilde{\Omega}_{\tilde{b}}(\vec{z}), G(\vec{x}')\}^*,
\end{split}
\end{equation}
where, as mentioned earlier, the brackets on the right-hand side are the Dirac brackets computed with respect to the first set. For the physical variables the nonvanishing brackets turn out to be
\begin{gather}
\label{bravsinc} \{s(\vec{x}), v_i(\vec{x}'\}^*_\mathrm{inc} = - \frac{\partial_i s}{\rho} \delta(\vec{x}-\vec{x}') + \frac{\partial_j s}{\rho} \partial_j \Delta^{-1}\partial_i \delta(\vec{x}-\vec{x}'), \\
\label{braBvinc}
\begin{split} \{B_i(\vec{x}), v_j(\vec{x}')\}^*_\mathrm{inc} &= \delta_{ij} \left( \frac{B_k}{\rho} \right)_{x'} \partial_k \delta(\vec{x}-\vec{x}') - \left( \frac{B_i}{\rho} \right)_{x'} \partial_j \delta(\vec{x}-\vec{x}') \\ &{}\quad + \partial_k\left[ \frac{1}{\rho}(B_i \partial_k \Delta^{-1}\partial_j \delta(\vec{x}-\vec{x}') - B_k \partial_i \Delta^{-1}\partial_j \delta(\vec{x}-\vec{x}'))\right],
\end{split} \\
\label{bravvinc}
\begin{split} \{v_i(\vec{x}), v_j(\vec{x}')\}^*_\mathrm{inc} &= \frac{\omega_{ij}}{\rho} \delta(\vec{x}-\vec{x}') + \partial_i \Delta^{-1}\partial_l \left[ \frac{1}{\rho}\omega_{lk}\partial_k \Delta^{-1}\partial_j \delta(\vec{x}-\vec{x}')\right] \\&{}\quad - (\omega_{kj}/\rho)_{x'} \partial_i\Delta^{-1}\partial_k \delta(\vec{x}-\vec{x}') - (\omega_{ik}/\rho) \partial_k\Delta^{-1}\partial_j \delta(\vec{x}-\vec{x}').
\end{split}
\end{gather}
The bracket of $\rho$ with $v_i$ now vanishes: $\{\rho(\vec{x}), v_i(\vec{x}')\}^*_\mathrm{inc} = 0$. Now implementing the constraints, the total Hamiltonian \eqref{hamtmhdinc2} reduces to the canonical Hamiltonian
\begin{equation}\label{hammhdinc2}
H^\mathrm{inc} = \int\!\ud^3 x \Big( \frac{1}{2} \rho v^2 + \rho \epsilon(\rho,s) + \frac{B^2}{2\mu} \Big),
\end{equation}
from which the equations of motion follow using the Dirac brackets \eqref{bravsinc}--\eqref{bravvinc}. Since $\rho$--$v_i$ bracket now vanishes it follows that $\partial_0 \rho = 0$, which is compatible with $\rho=\rho_0$. Other equations of motion following from \eqref{hammhdinc2} and \eqref{bravsinc}--\eqref{bravvinc} are
\begin{gather}
\label{eqsinc2} \partial_0 s  = - v_i \partial_i s + \partial_i s \partial_i \Delta^{-1} (\partial_j v_j), \\
\label{eqBinc2} \partial_0 B_i = B_j \partial_jv_i - v_j \partial_j B_i + \partial_j B_i \partial_j \Delta^{-1}(\partial_k v_k) - B_j \partial_j \partial_i \Delta^{-1}(\partial_k v_k), \\
\label{eqvinc2}
\begin{split} \partial_0 v_i &= v_j \omega_{ij} + \frac{1}{\mu \rho_0} B_j (\partial_j B_i - \partial_i B_j) - \omega_{ik} \partial_k \Delta^{-1} (\partial_j v_j) + \partial_i\Delta^{-1}\partial_l[\omega_{lk}\partial_k \Delta^{-1}(\partial_j v_j)] \\
&{}\qquad -\partial_i \Delta^{-1}\partial_k (v_j \omega_{kj}) + \frac{1}{\mu\rho_0}\partial_i\Delta^{-1}\partial_k (B_j\partial_k B_j - B_j \partial_j B_k),
\end{split}
\end{gather}
which using $\partial_i v_i = 0$ simplify to
\begin{gather}
\label{eqsinc22} \partial_0 s  = - v_i \partial_i s , \\
\label{eqBinc22} \partial_0 B_i = B_j \partial_jv_i - v_j \partial_j B_i, \\
\label{eqvinc22} \begin{split}\partial_0 v_i &= v_j \omega_{ij} + \frac{1}{\mu \rho_0} B_j (\partial_j B_i - \partial_i B_j)\\ &{}\quad - \partial_i \left[ \Delta^{-1}\partial_k (v_j \omega_{kj}) - \frac{1}{\mu\rho_0}\Delta^{-1}\partial_k (B_j\partial_k B_j - B_j \partial_j B_k) \right].\end{split}
\end{gather}
Equation \eqref{eqvinc22} appears to be different from \eqref{eqvinc} but is actually not. A few mathematical manipulations and the use of \eqref{new3} on the right-hand side of \eqref{eqvinc22} reduces it to \eqref{eqvinc}. Thus, the two sets \eqref{eqsinc}--\eqref{eqvinc} and \eqref{eqsinc22}--\eqref{eqvinc22} match. This shows that both the descriptions are consistent.

The use of Dirac brackets for reduction of general (compressible) MHD to incompressible MHD has earlier been discussed in \cite{CMT2012, CGBTM2013, Chandre2015}. Equations \eqref{eqsinc22}--\eqref{eqvinc22} are identical to those of \cite{CMT2012}. In such studies, however, one has to already start from the noncanonical brackets of general (compressible) MHD and the role of Dirac brackets is to confine to the case of incompressible MHD. Here we have demonstrated how one can obtain the noncanonical (Dirac) brackets for incompressible MHD by incorporating the condition of incompressibility in the action of general MHD and following Dirac's method. We also presented yet another approach, which is completely new, in terms of standard Poisson brackets but involving a complicated Hamiltonian, the total Hamiltonian \eqref{hamtmhdinc2}, to obtain the equations of motion for incompressible MHD.

%%%%%%%%%%%%%%%%%%%%%%%%%%%%%%%%%%%%%%%%%%%%%%%%%%%%%%%%%%%%%%%%%%

\section{\label{mhdemt}Energy-momentum tensor}

For a field coupled to an external electromagnetic field, the energy-momentum tensor \eqref{emt} satisfies\footnote{Our convention: $F^{0i}=E_i$, $F^{12}=B_3$, $F^{23}=B_1$, $F^{31}=B_2$.}
\begin{equation}\label{dtfj}
\partial _\kappa T^{\kappa \nu} = {F^\nu}_\kappa J^\kappa.
\end{equation}
Let us first consider the $\nu=0$ part of \eqref{dtfj}:
\begin{equation}\label{dtfj0}
\partial _\kappa T^{\kappa 0} = {F^0}_\kappa J^\kappa.
\end{equation}
Evaluating the right-hand side of \eqref{dtfj0} for our theory yields
\begin{equation}
\text{RHS of \eqref{dtfj0}} = {F^0}_\kappa J^\kappa = E_i (\rho v_i) = (-\varepsilon_{ikm}v_k B_m) (\rho v_i) = 0,
\end{equation}
where we have used the infinite-conductivity version of Ohm's law, $\vec{E} = -\vec{v} \times \vec{B}$, to eliminate $E_i$. Similarly, for $\nu=j$, \eqref{dtfj} reads
\begin{equation}\label{dtfjj}
\partial _\kappa T^{\kappa j} = {F^j}_\kappa J^\kappa,
\end{equation}
and we again find, using $\vec{E} = -\vec{v} \times \vec{B}$,
\begin{equation}\begin{split}
\text{RHS of \eqref{dtfjj}} &= F^{j0}J_0 + F^{jk}J_k=-E_j(-\rho)+\varepsilon_{jkm}B_m(\rho v_k)\\ &= \varepsilon_{jkm}v_k B_m(-\rho)+\varepsilon_{jkm}B_m(\rho v_k) = 0.
\end{split}
\end{equation}

Now we shall do an explicit computation of the left-hand sides of \eqref{dtfj0} and \eqref{dtfjj} to check consistency of our analysis. For the Lagrangian given in \eqref{actmhd} we find
\begin{gather}
T^{00} = - \mathcal{H}= -\Big( \frac{1}{2} \rho v^2 + \rho \epsilon(\rho,s) + \frac{B^2}{2\mu} \Big), \\
T^{0i} = -\left( \theta\partial_i \rho + \lambda\rho\partial_i s + \alpha\rho \partial_i \beta + K_j \partial_i B_j \right), \\
T^{i0} = -\rho v_i \partial_0 \theta + \rho\lambda v_i \partial_0 s + \rho\alpha v_i \partial_0 \beta + B_i v_j \partial_0 K_i - B_j v_i \partial_0 K_j, \\
T^{ij} = -\rho v_i v_j -\delta_{ij}\mathcal{L} + v_i B_m \partial_m K_j - v_m B_i \partial_j K_m,
\end{gather}
where $v_i$ is as given in \eqref{eqv1}. From these expressions we now compute $\partial_0 T^{00}$ and $\partial_i T^{i0}$ and make use of the equations of motion \eqref{eqrho}--\eqref{eqB}, \eqref{eqbeta}--\eqref{eqtheta} and \eqref{eqK}. This yields
\begin{equation}
\text{LHS of \eqref{dtfj0}} = \partial _\kappa T^{\kappa 0} = \partial_0 T^{00} + \partial_i T^{i0} = 0.
\end{equation}
In similar manner, we compute $\partial_0 T^{0j}$ and $\partial_i T^{ij}$, and after somewhat lengthy algebra we get
\begin{equation}
\text{LHS of \eqref{dtfjj}} = \partial _\kappa T^{\kappa j} = \partial_0 T^{0j} + \partial_i T^{ij} = 0.
\end{equation}
This verifies \eqref{dtfj} and expresses the conservation of energy and momentum in nonrelativistic ideal MHD. Thus, the interaction of fluid ``particles'' with the magnetic field is localised; had the fluid ``particles'' been subject to forces that act at a distance, the energy-momentum tensor would not be conserved \cite{W1972}.

%%%%%%%%%%%%%%%%%%%%%%%%%%%%%%%%%%%%%%%%%%%%%%%%%%%%%%%%%%%%%%%%%%

\section{\label{conclu}Summary and conclusion}

We have given a new Lagrangian in Euler variables for an ideal nonrelativistic MHD extending an earlier approach \cite{BM2015}, involving one of us, in the context of fluid dynamics. A distinctive feature of that approach, briefly reviewed in Sec.~\ref{mhdact}, was the natural appearance of Clebsch form for the fluid velocity. These ideas were extended to include MHD, leading to a generalised Clebsch decomposition involving the magnetic field. The final Lagrangian involves 12 basic fields including the physical ones, like fluid density, entropy density and magnetic field. It was shown to yield the MHD equations from a variational approach. We also discussed the connection of this Lagrangian with that given in \cite{BO2000}.

To obtain a Hamiltonian formulation we followed Dirac's constraint analysis. The MHD system turned out to be second-class since all constraints appearing there were shown to be second-class. These constraints were eliminated by the construction of Dirac brackets. These brackets were just the noncanonical brackets first posited in \cite{MG1980}. Since Dirac brackets, which manifestly satisfy the Jacobi identity, have been used, it is not necessary to check this property, as in other approaches where noncanonical brackets have been discussed. The MHD equations were obtained by using these brackets and the standard canonical Hamiltonian. A complementary viewpoint within this formulation was also discussed. There the constraints were not eliminated. Rather, they were implemented by Lagrange multipliers in the construction of the total Hamiltonian. The multipliers were obtained by requiring time-conservation of the constraints. In this interpretation, all the brackets were canonical. The MHD equations were once again rederived, now from this (total) Hamiltonian but with canonical brackets.

It is pertinent to note that the Hamiltonian formulation involving noncanonical brackets already exists in the literature. To arrive at these brackets either recourse has to be taken by considering the Lagrange version first and using its mapping to the Euler version or choosing a suitable Clebsch version of Hamiltonian, identifying the canonical pairs of variables and then mapping to the noncanonical brackets for physical variables. We have provided a systematic way, starting from an action principle and using Dirac's constraint analysis, to obtain these noncanonical brackets solely using Euler variables. We have also shown that it is possible to describe MHD with canonical brackets in an enlarged phase space by including the momenta conjugate to all the variables appearing in the action. It is just a trading between the canonical Hamiltonian having noncanonical brackets and a noncanonical Hamiltonian possessing canonical brackets. Thus we reveal a richer structure of MHD than has been reported earlier. Our approach to obtain the Hamiltonian formulation of MHD based on Dirac's method is completely new.

We also considered the case of incompressible MHD. We incorporated the condition of incompressibility in the action itself and obtained the Hamiltonian formulation following Dirac's method. This was in contrast with the earlier studies \cite{CMT2012, CGBTM2013, Chandre2015} where one has to already start from the noncanonical brackets of general (compressible) MHD and the role of Dirac brackets is just to confine to the case of incompressible MHD. We started from an action principle, the constraints of the theory followed naturally from the Lagrangian and the definitions of canonical momenta, and finally Dirac brackets were obtained. Moreover, we have shown that it is possible to describe incompressible MHD also in terms of the standard canonical Poisson brackets and a modified (noncanonical) Hamiltonian.

For the consistency of our analysis we explicitly computed the energy-momentum tensor following Noether's definition and showed its conservation. The conservation of the MHD stress tensor is significant. It implies that the interactions of the charged fluid are local and confined to a short range \cite{W1972}. This property would be destroyed if long-range interactions were present.

As a final remark, we would like to apply this analysis to the Hall and extended MHD. These models are extended versions of MHD. Their brackets were posited in \cite{YH2013, AKY2015} and very recently \cite{DML2015} derived through the Lagrange to Euler map.

%%%%%%%%%%%%%%%%%%%%%%%%%%%%%%%%%%%%%%%%%%%%%%%%%%%%%%%%%%%%%%%%%%

\section*{Acknowledgement}

KK would like to acknowledge the financial support from S.N. Bose National Centre for Basic Sciences, Kolkata, where this work was carried out.

%%%%%%%%%%%%%%%%%%%%%%%%%%%%%%%%%%%%%%%%%%%%%%%%%%%%%%%%%%%%%%%%%%

%%%%%%%%%%%%%%%%%%%%%%%%%%%%%%%%%%%%%%%%%%%%%%%%%%%%%%%%%%%%%%%%%%

\end{document}